\documentclass[]{aa} % for the letters 
\usepackage{txfonts,epsfig,graphicx,natbib,url,twoopt}

\usepackage{natbib,twoopt}
\usepackage[breaklinks=true]{hyperref} %% to avoid \citeads line fills
\bibpunct{(}{)}{;}{a}{}{,}             %% natbib format for A&A and ApJ
\makeatletter
  \newcommandtwoopt{\citeads}[3][][]{\href{http://adsabs.harvard.edu/abs/#3}%
    {\def\hyper@linkstart##1##2{}%
     \let\hyper@linkend\@empty\citealp[#1][#2]{#3}}}
  \newcommandtwoopt{\citepads}[3][][]{\href{http://adsabs.harvard.edu/abs/#3}%
    {\def\hyper@linkstart##1##2{}%
     \let\hyper@linkend\@empty\citep[#1][#2]{#3}}}
  \newcommandtwoopt{\citetads}[3][][]{\href{http://adsabs.harvard.edu/abs/#3}%
    {\def\hyper@linkstart##1##2{}%
     \let\hyper@linkend\@empty\citet[#1][#2]{#3}}}
  \newcommandtwoopt{\citeyearads}[3][][]%
    {\href{http://adsabs.harvard.edu/abs/#3}
    {\def\hyper@linkstart##1##2{}%
     \let\hyper@linkend\@empty\citeyear[#1][#2]{#3}}}
\makeatother

\begin{document}

   \title{Estimating the magnetic field strength from magnetograms}

   \author{A. Asensio Ramos \and M. J. Mart\'{\i}nez Gonz\'alez \and R. Manso Sainz}

   \institute{Instituto de Astrof\'\i sica de Canarias,
              38205, La Laguna, Tenerife, Spain; \email{aasensio@iac.es}
            \and
Departamento de Astrof\'{\i}sica, Universidad de La Laguna, E-38205 La Laguna, Tenerife, Spain
             }
             
  \date{Received ---; accepted ---} 

  \abstract{A properly calibrated longitudinal magnetograph is an instrument that measures circular polarization and gives an estimation of the magnetic flux density in each 
  observed resolution element. This usually constitutes a lower bound of the field strength in the resolution element, given that it can be made
  arbitrarily large as long as it occupies a proportionally smaller area of the resolution element and/or becomes more transversal to the
  observer and still produce the same magnetic signal. Yet, we know that \emph{arbitrarily stronger} fields are less likely --hG fields are more probable than kG fields, with
  fields above several kG virtually absent-- and we may even have partial information about its angular distribution.
  Based on a set of sensible considerations, we derive simple formulae based on a Bayesian analysis to give an improved estimation of the magnetic field strength
  for magnetographs.}

   \keywords{Sun: magnetic fields, atmosphere --- line: profiles --- methods: statistical, data analysis}
   \authorrunning{Asensio Ramos et al.}
   \titlerunning{Estimating the magnetic field strength from magnetograms}
   \maketitle
%
%________________________________________________________________

\section{Introduction}
The activity of the Sun (and of a large percentage of other stars) is driven by the presence of magnetic fields.
It was realized long ago that it is important to carry out systematic observations of the surface magnetism of the
Sun if we ever want to understand its activity cycle and that of other stars. This possibility become a reality
after the discovery of the Zeeman effect \citep{zeeman1897} that a magnetic field produces on some spectral lines (leading to a splitting in frequency of the $\sigma$ and $\pi$ components),
and its discovery on the Sun by \cite{hale1908}. 

% The usage of slit spectrographs was not convenient for such synoptic observations given the impossibility of scanning the full solar disk.
A breakthrough occured when \cite{babcock53} developed the longitudinal magnetograph, a device that measured the line-of-sight (LOS) component of the magnetic field over
an area of the surface of the Sun.
In order to measure the magnetic field, such apparatus performs a measurement of the intensity $I(\lambda)$ and circular polarization $V(\lambda)$
of a magnetically sensitive spectral line in a certain narrow filter with transmission $p(\lambda)$, and returns the following quantity \citep[e.g.,][]{landi_landolfi04}:
\begin{equation}
S_V = \frac{\int \mathrm{d}\lambda\, V(\lambda) p(\lambda)}{\int \mathrm{d}\lambda\, I(\lambda) p(\lambda)}.
\end{equation}
The importance of this quantity resides on the fact that it can be easily related to the magnetic flux density ($\Phi$) across the resolution element
of the instrument (pixel)\footnote{Although the units Mx cm$^{-2}$ and G are equivalent, we follow \cite{keller_94} and use the former for the magnetic flux density 
to make its observational character explicit.}. To this end, 
a number of simplifying conditions have to be imposed, the most important of which is the assumption that
the field is in the weak-field approximation \citep[e.g.,][]{landi_landolfi04}.
% In other words, the Zeeman splitting has to be smaller than 
% the intrinsic broadening of the line . 
If this holds, the measured signal
is simply given by
\begin{equation}
S_V = C \Phi,
\end{equation}
where $C$ is a calibration constant that needs to be obtained from a proper modeling of the line formation
mechanism and depends on the exact details of the filter $p(\lambda)$ and the selected spectral line. Once the instrument
is calibrated, $\Phi=S_V/C$ gives the estimated magnetic flux density. When the calibration is assumed to be perfectly
known and the noise in $S_V$ is Gaussian, $\Phi=S_V/C$ constitutes the unbiased maximum likelihood estimation of the magnetic flux
density.

From its conception, many longitudinal magnetographs have been on routine operation around the globe and on
space. One of the most used is the Michelson Doppler Imaging (MDI) onboard the Solar and Heliospheric Observatory (SoHO),
which provided synoptic maps of magnetic flux density. 
% Another recent successful linear magnetograph was the
% IMaX instrument \citep{imax11}. Although it was a vector magnetograph, it was also provided with a purely linear
% magnetograph mode.
Obviously, the fundamental drawback of any longitudinal magnetograph is that it only gives us partial information about
the magnetic field in the observed region \citep{lites99}, giving a lower limit to the strength.
% if the noise is not too large.
% As we show later, in the simplest case, the signal measured by the
% instrument is proportional to the line-of-sight (LOS) component of the magnetic field. 
We propose in this note, using simple Bayesian inference, a way to probabilistically correct the measurements of
longitudinal magnetographs to provide a better estimation of the magnetic field strength, instead of just a lower limit. The correction
relies on some sensible statistical a-priori assumptions about the properties of the field in the observed resolution element
that can be modified at will. This facilitates the adaptation of our suggested correction to any
desired a priori assumption imposed by the researcher.

%%%%%%%%%%%%%%%%%%%%%%%%%%%%%%%%%%%%%%%%%%%%
\section{Bayesian analysis}
\subsection{Generative models}
The first step in any Bayesian analysis is to write down an expression that
describes how the observations are generated, the so-called generative model.
If the field in the observed resolution element is resolved at the spatial resolution of the instrument (i.e., the magnetic field is assumed to 
be homogeneous and filling the whole resolution element), then the magnetic flux density can be directly related
to the magnetic field strength $B$ in the weak-field limit \citep[e.g.,][]{landi_landolfi04}. Without loss of generality, we assume that
$C=1$, so that our measurement is $\Phi=S_V$. This avoids carrying the instrument-specific constant $C$ in all
calculations. Therefore, the generative model for the resolved case is
\begin{equation}
\Phi_\mathrm{r} = B \mu + \epsilon,
\end{equation}
where $\mu=\cos \theta$ is the cosine of the angle between the LOS and the magnetic field vector, while
$\epsilon$ stands for the measurement noise, which we assume to have Gaussian statistics with variance $\sigma_n^2$. Only measuring
$\Phi_\mathrm{r}$, a degeneracy between $B$ and $\mu$ appears that cannot be resolved unless more information
is added. We show in the following that the inclusion of priors helps partially resolving this degeneracy.

When the field in the resolution element of our instrument is not resolved, it is useful to introduce the concept 
of a filling factor, $f\in [0,1]$. This quantity accounts for the fraction
of the resolution element that is filled with an homogeneous magnetic field. The remaining $1-f$ fraction
is assumed to be field-free.
Under the assumption that the presence of a magnetic field does not modify the thermodynamic conditions, the generative model for the
non-resolved case is
\begin{equation}
\Phi_\mathrm{nr} = f B \mu + \epsilon.
\label{eq:generative_nonresolved}
\end{equation}
It is possible to generalize these formulae for the case in which the thermodynamic conditions are different
in the magnetic and the non-magnetic regions \citep{landi_landolfi04}, but we prefer to stick to these simple cases. Our
analysis is fully generalizable to these cases, though.

% In the more complex case that the intensity profile in the magnetic and the non-magnetic region of the pixel are not
% equal, the generative model becomes more difficult. However, if we assume that the intensity profile in the
% magnetic region is just a scaling of the intensity profile in the non-magnetic region, so that 
% $I_\mathrm{m}(\lambda) = \Theta I_\mathrm{nm}(\lambda)$, then the generative model simplifies to \citep{landi_landolfi04}
% \begin{equation}
% \Phi_\mathrm{nr2} = \frac{f \Theta}{1-f+f\Theta} B \mu + \epsilon.
% \label{eq:generative_nonresolved_theta}
% \end{equation}

\subsection{Posterior distribution}
Once the magnetic flux density $\Phi_\mathrm{obs}$ is measured, all the information about the model parameters of interest is
encoded in the posterior distribution. The posterior distribution for the generative model of Eq. (\ref{eq:generative_nonresolved})
is, applying the Bayes rule, given by
\begin{equation}
p(B,\mu,f|\Phi_\mathrm{obs}) = \frac{p(\Phi_\mathrm{obs}|B,\mu,f) p(B,\mu,f)}{p(\Phi_\mathrm{obs})},
\label{eq:fullPosterior}
\end{equation}
where $p(B,\mu,f|\Phi_\mathrm{obs})$ is the posterior for the model parameters, $p(\Phi_\mathrm{obs}|B,\mu,f)$ is the likelihood
that encodes the ability of the model to reproduce the observations, $p(B,\mu,f)$ is the prior distribution
that summarizes all the a-priori information we have about the parameters. Finally, $p(\Phi_\mathrm{obs})$ is the evidence, a normalization
constant that is unimportant here because it does not depend on the model parameters and that we drop in the following.

According to the statistics of the noise proposed in the generative model, the likelihood is given by a Gaussian distribution with variance $\sigma_n^2$:
\begin{equation}
\mathcal{L} = p(\Phi_\mathrm{obs}|B,\mu,f) = \frac{1}{\sqrt{2 \pi}\sigma_n} \exp \left[-\frac{\left( \Phi_\mathrm{obs}-Bf\mu \right)^2}{2\sigma_n^2} \right].
\end{equation}
For the sake of simplicity, we assume that $B$, $\mu$ and $f$ are a-priori statistically independent, so that the prior factorizes:
\begin{equation}
p(B,\mu,f) = p(B) p(\mu) p(f).
\end{equation}
Note that this step is just a simplification and more complicated a-priori information can be easily introduced.
In particular, it is possible to easily introduce the fact that stronger fields tend to be
more vertical by writing a combined prior $p(B,\mu)$ that takes this property into account.

Our state of knowledge about the magnetic field strength after carrying out the observation with the
magnetograph is computed by integrating $f$ and $\mu$ from the posterior distribution of Eq. (\ref{eq:fullPosterior}). 
Assuming complete ignorance on $f$ by using a flat prior on the interval $[0,1]$, the
ensuing marginal posterior is:
\begin{align}
p(B|\Phi_\mathrm{obs}) &\propto p(B) \int_0^1 \mathrm{d}f p(f) \int_{-1}^1 \mathrm{d}\mu p(\mu) p(\Phi_\mathrm{obs}|B,\mu,f) \nonumber \\
&= p(B) \int_{-1}^1 \mathrm{d}\mu p(\mu) \frac{1}{B\mu}\left[ \mathrm{erf}\left( \frac{\Phi_\mathrm{obs}}{\sqrt{2} \sigma_n} \right) - \mathrm{erf}\left( \frac{\Phi_\mathrm{obs} - B\mu}{\sqrt{2} \sigma_n} \right)\right]
\label{eq:posterior}
\end{align}
Once the posterior distribution is computed, it is easy to estimate upper limits to the magnetic field strength at a certain
credibility limit. We later calculate B$_{68}$ and B$_{95}$, as the 68\% and 95\% upper limits.

It is key to understand that the posterior distribution for the magnetic field strength will change
depending on the election of the specific prior. Although this is obviously always the case, it is of special relevance
in our problem because of the presence of strong degeneracies. For this reason, it is advisable to consider 
several priors and see how the results change with them. The first prior we consider is a log-normal distribution, 
as pointed out by \cite{vogler_thesis03} and \cite{vogler05} using magnetoconvection simulations:
\begin{equation}
p(B) = \frac{1}{B \sigma_b \sqrt{2 \pi}} \exp \left[ - \frac{(\ln B - \ln B_0)^2}{2\sigma_b^2}\right],
\label{eq:lognormal_prior}
\end{equation}
where $B_0$ and $\sigma_b$ are free hyperparameters, that we choose to be $B_0=38$ G and $\sigma_b=1.2$, following the results of
\citep{dominguez06b}. Note that the average of a log-normal random variable equals $\exp(\ln B_0+\sigma_b^2/2)$, which equals 78 G for the
values of the parameters. Although this value is slightly below those estimated by \cite{trujillo_nature04}, \cite{shchukina_trujillo11} and \cite{rempel14},
it decreases slower towards large magnetic field strengths than the exponential distribution present in the simulations or assumed by \cite{trujillo_nature04}.
Our second prior is an exponential distribution
\begin{equation}
p(B) = \frac{1}{B_0} \exp\left(-\frac{B}{B_0} \right),
\label{eq:exponential_prior}
\end{equation}
with $B_0=130$ G \citep{trujillo_nature04,rempel14} or $B_0=60$ G \citep{rempel14}. We advance that, although the three considered priors are 
very different, the results are relatively robust to this election.

\begin{figure*}
\centering
\includegraphics[width=\textwidth]{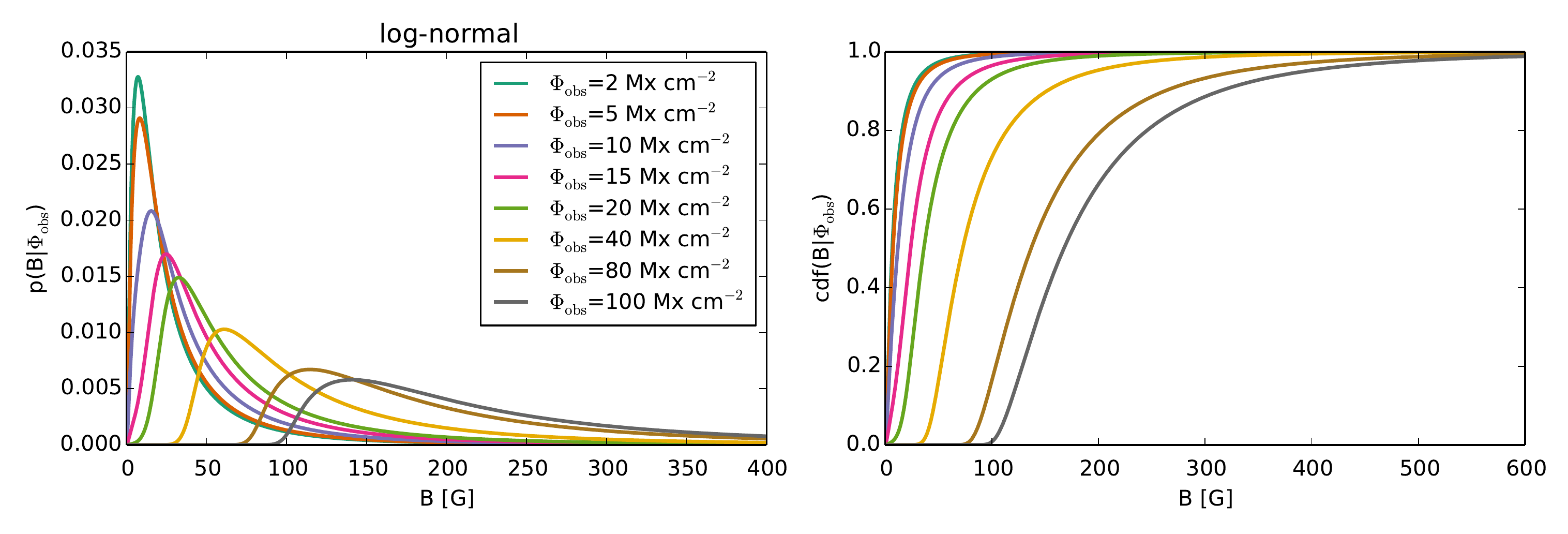}
\includegraphics[width=\textwidth]{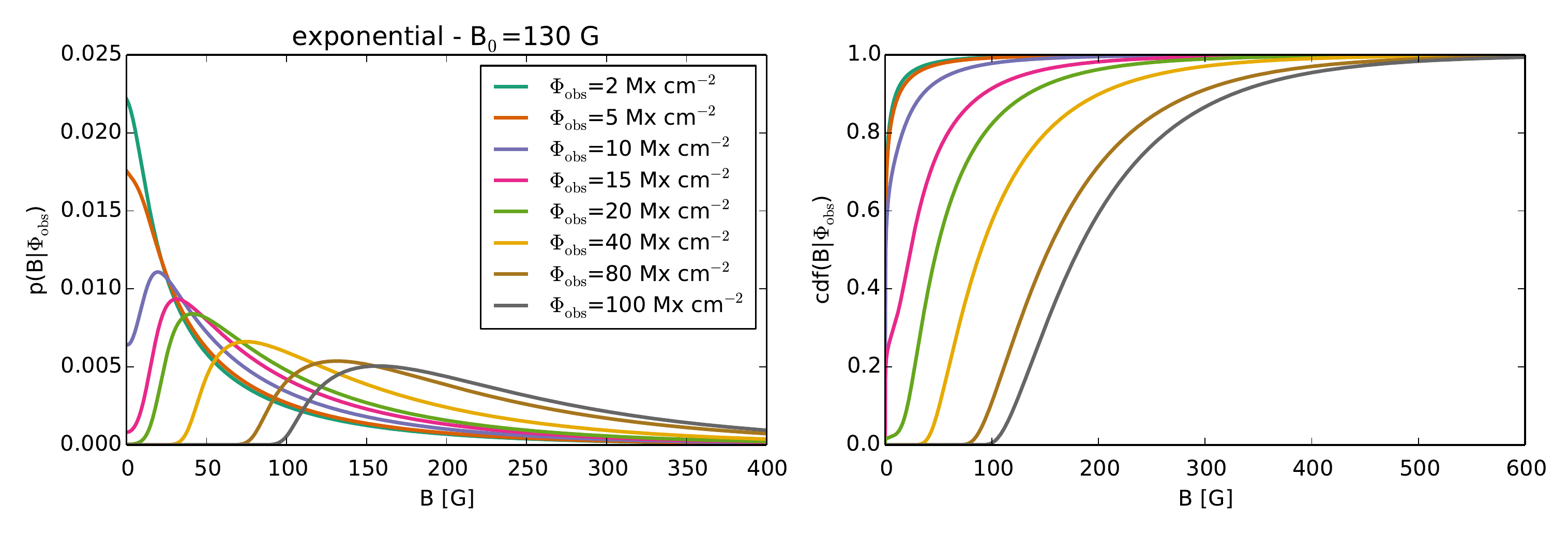}
\includegraphics[width=\textwidth]{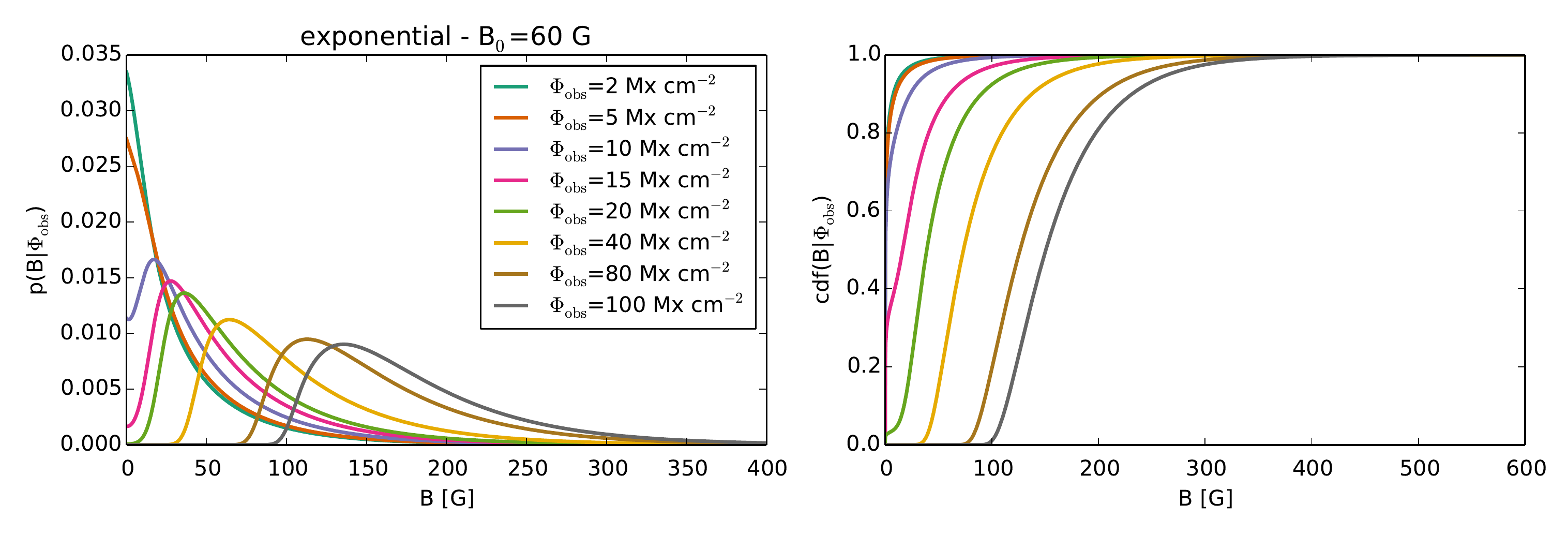}
\caption{Marginal posterior distributions from Eq. (\ref{eq:posterior}) (left panel) and their ensuing cumulative distributions (right panel) for 
a standard deviation of the noise of $\sigma_n=5$ Mx cm$^{-2}$ and different values of the observed
magnetic flux density. The actual values of $\Phi_\mathrm{obs}$ are indicated in the legend. Flat priors for $f$ and $\mu$ are used. Each
row shows the calculations for the different priors: (top) a log-normal prior with $B_0=38$ G and $\sigma_b=1.2$, (middle) an exponential
prior with $B_0=130$ G, and (bottom) an exponential prior with $B_0=60$ G.}
\label{fig:posteriorFlux}
\end{figure*}

\section{Results}
\subsection{Ignorance about the topology of the field}
\begin{figure*}
\centering
\includegraphics[width=\textwidth]{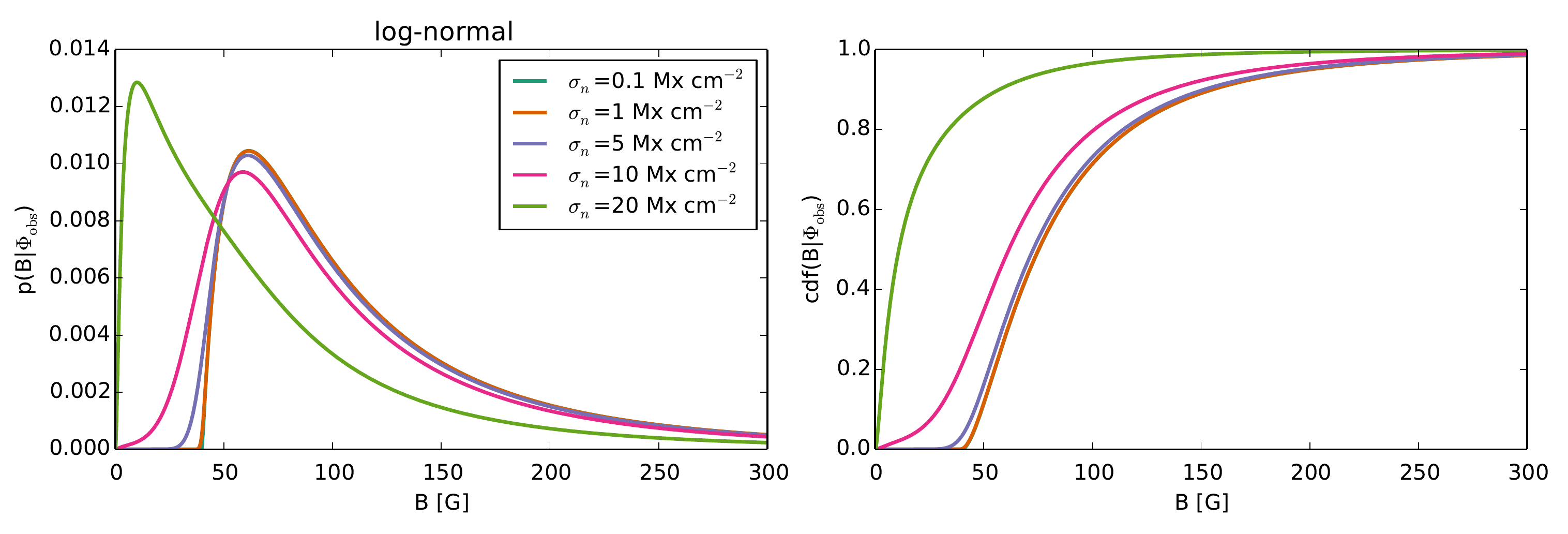}
\includegraphics[width=\textwidth]{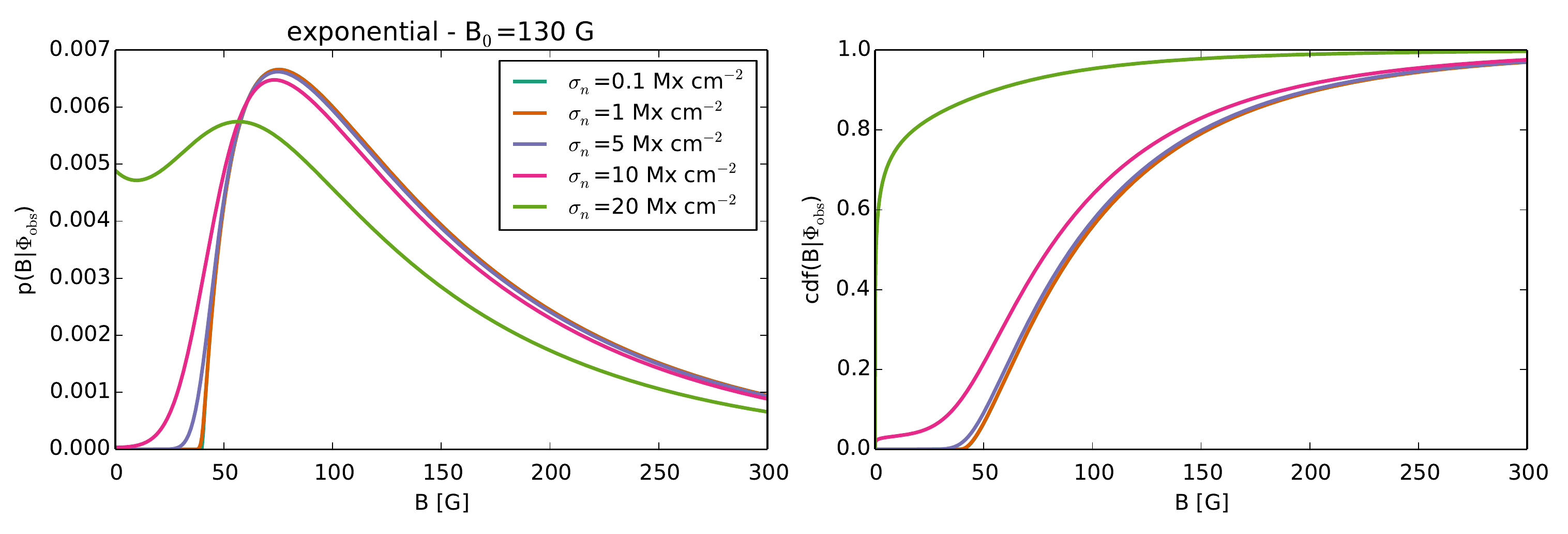}
\includegraphics[width=\textwidth]{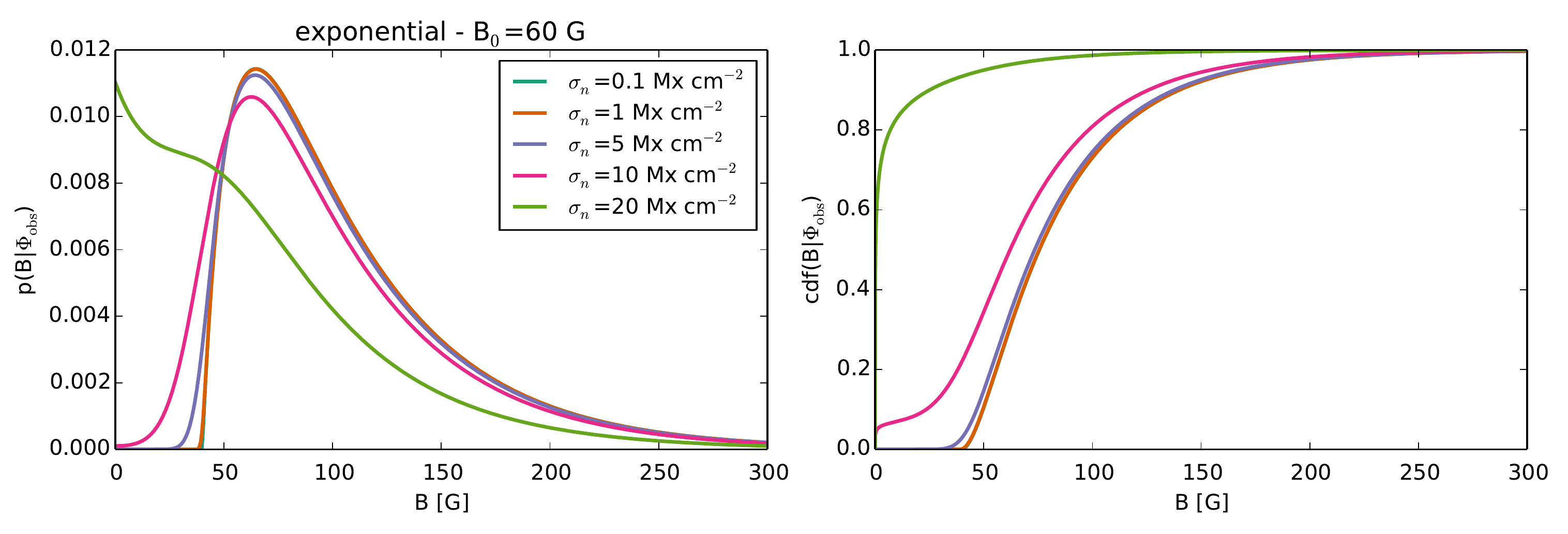}
\caption{Same as Fig. \ref{fig:posteriorFlux} for a fixed observed magnetic flux of $\Phi_\mathrm{obs}=40$ Mx cm$^{-2}$ and
different values of the standard deviation of the noise $\sigma_n$.}
\label{fig:posteriorNoise}
\end{figure*}

When we are completely ignorant about the topology of the field, a reasonable assumption is to 
consider it, a-priori, isotropic. To this end, we consider a flat distribution in $\mu$, which 
yields a vector field uniformly distributed in solid angle.
Figure \ref{fig:posteriorFlux} displays the marginal posteriors for a fixed standard deviation of the
noise of $\sigma_n=5$ Mx cm$^{-2}$, typical of modern longitudinal magnetographs\footnote{The code to reproduce the figures can be
found in \texttt{https://github.com/aasensio/magnetographCorrection}.}. We show the marginal posteriors 
for different values of the observed magnetic flux density.
The left panels display the marginal posterior, while the right panels show the cumulative distribution.  
The computation is done using a proper numerical quadrature in $\mu$. Each row corresponds to different
priors, as discussed above. We note that the effect of the specific prior is somehow marginal, except
in the tails of the posteriors. It is obvious that the exponential prior with $B_0=60$ G disfavors strong
fields and the tails of the posterior become less important, 

The curves of Fig. \ref{fig:posteriorFlux} are conditioned
on the actual measured value of $\Phi_\mathrm{obs}$. Consequently, consecutive measurements of exactly the same resolution element
would lead to slightly different curves.
In order to see the influence of the noise, we show in Fig. \ref{fig:posteriorNoise} the marginal posteriors for a fixed
value of $\Phi_\mathrm{obs}=40$ Mx cm$^{-2}$ and for different values of $\sigma_n$ for the three considered priors. The figure demonstrates that, when the
noise level is small, values of $B$ below $\Phi_\mathrm{obs}$ have very small probability. When the standard deviation of the noise
increases, values of $B<\Phi_\mathrm{obs}$ gain some probability. This result clearly exposes that, when the standard
deviation of the noise is of the order of the measured magnetic flux density, $\Phi_\mathrm{obs}$ is not anymore a lower bound of the
magnetic field strength. The differences among the priors are not very large except in the tails. This shows the robustness of the
inference to the specific chosen prior.

\begin{figure*}
\centering
\includegraphics[width=0.33\textwidth]{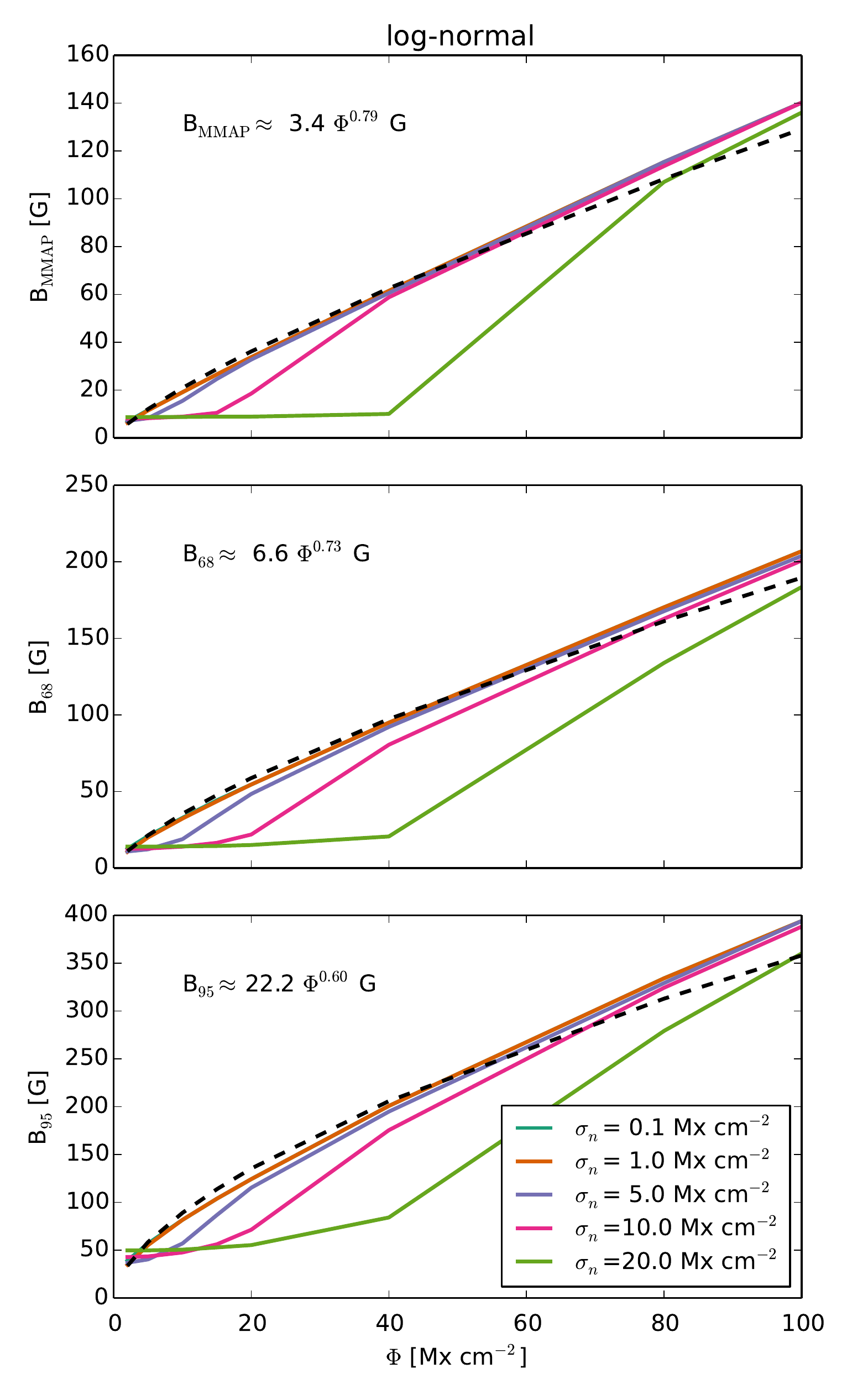}
\includegraphics[width=0.33\textwidth]{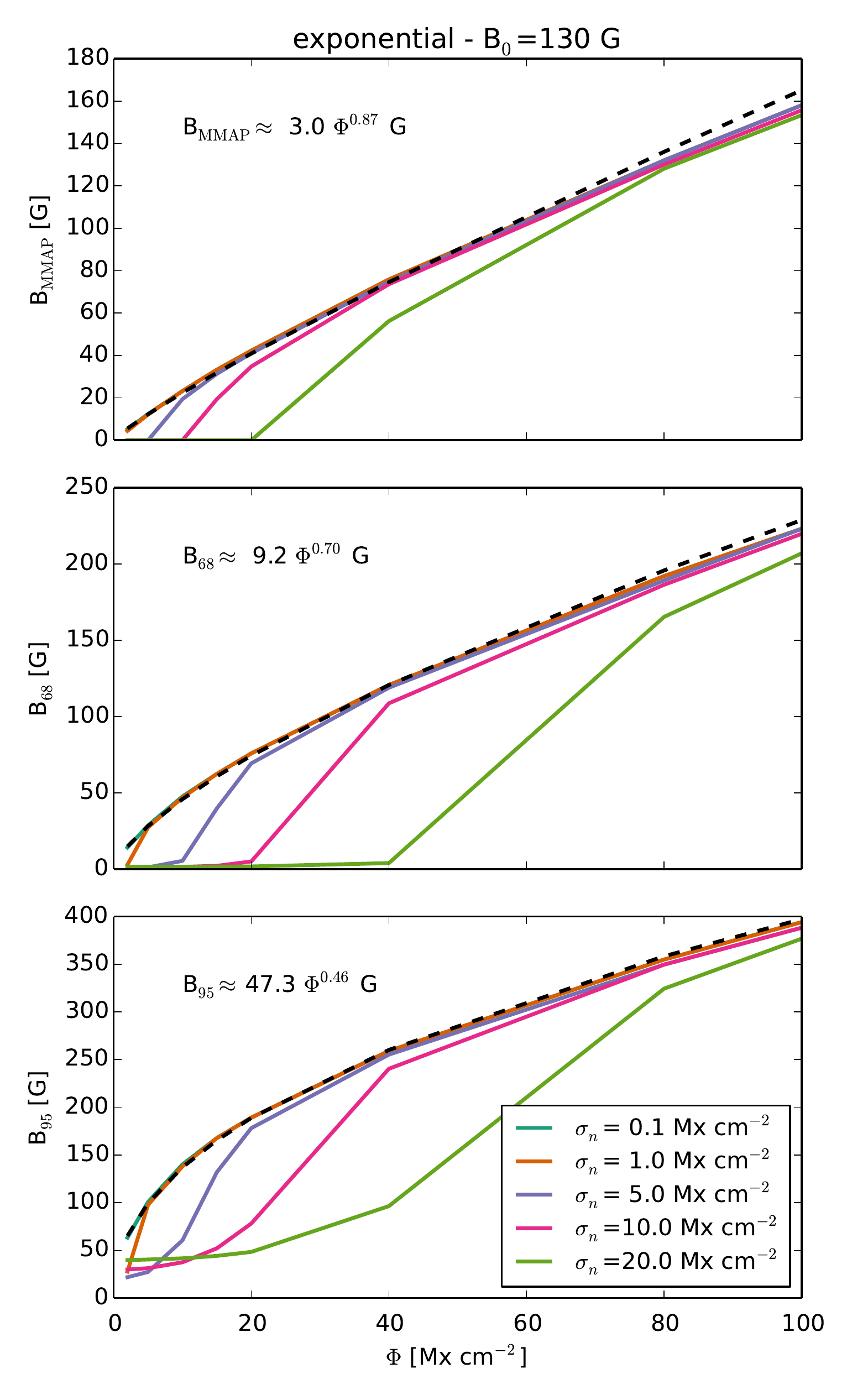}
\includegraphics[width=0.33\textwidth]{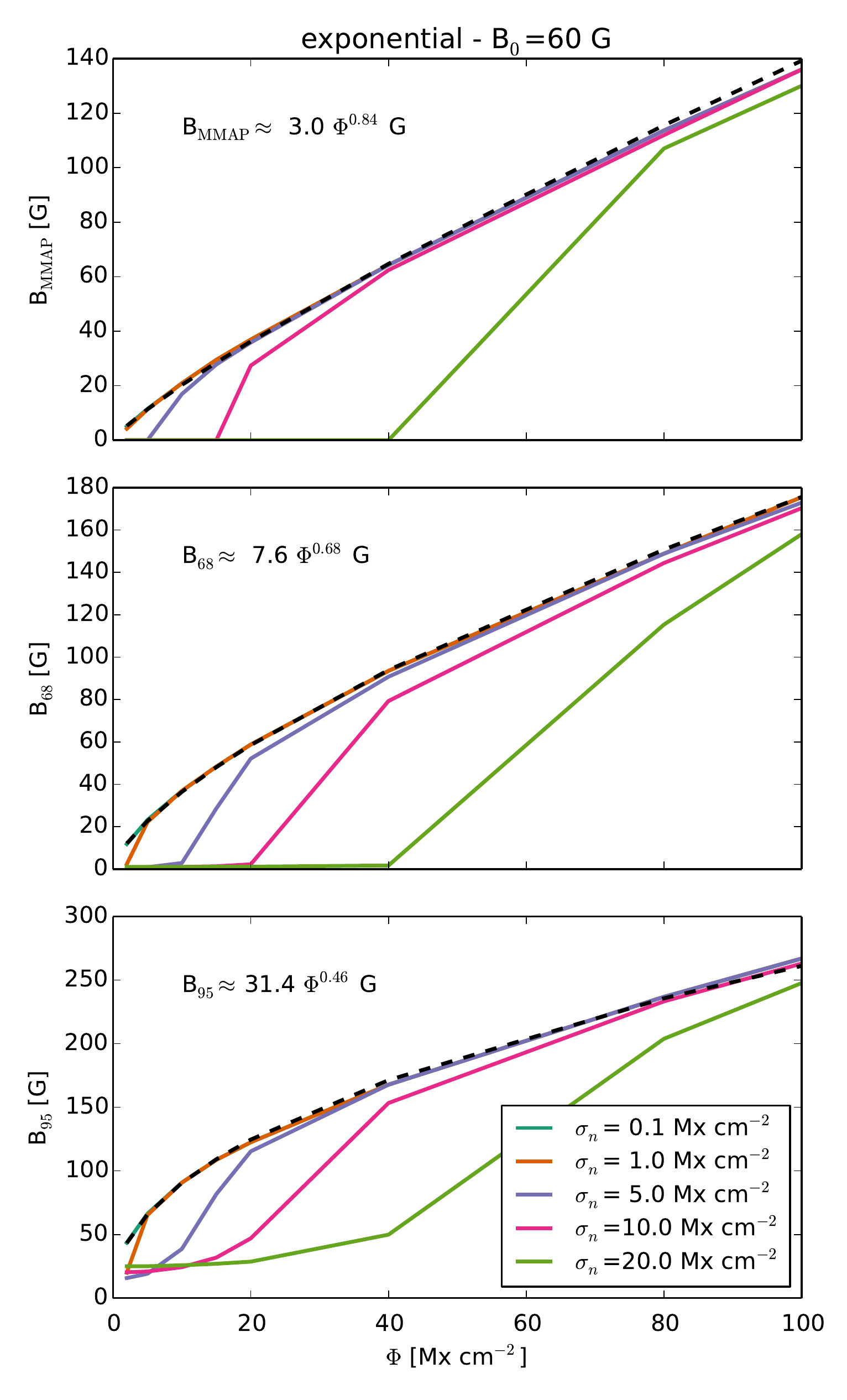}
\caption{Marginal maximum a-posteriori (upper panel), percentile 68 (middle panel) and 95 (lower panel) obtained
from the marginal posteriors computed from Eq. (\ref{eq:posterior}). Each curve corresponds to a different value of the
standard deviation of the noise. The dashed lines are the fits to the curves, a straight line for $B_\mathrm{MMAP}$
and a function of the type $\alpha \Phi^\beta$ for $B_{68}$ and $B_{95}$.}
\label{fig:calibration}
\end{figure*}

The marginal posteriors indicate that there is some available information about the magnetic field strength on a
single measure of the magnetic flux density. 
In essence, small values of $B$ are preferred with respect to large values. The peak of the marginal posterior 
$p(B|\Phi_\mathrm{obs})$ (the so-called marginal maximum a-posteriori [MMAP] value) 
for the magnetic field is shifted towards increasingly larger values
when the measured $\Phi_\mathrm{obs}$ increases.
We extracted the MMAP values of the magnetic field strength from these curves and we show them in the upper panels of 
Figure \ref{fig:calibration} with respect to the measured magnetic flux density for different values of the noise and for the
three considered priors. For the cases with standard deviations of the noise below 10 Mx cm$^{-2}$, it follows a relatively linear trend. For the remaining ones, large
variations are a consequence of the specific noise realization in the displayed posterior (it will change from realization to realization). We have displayed
a power-law fit to the curves. Our proposal is to use this value as an estimation of the most probable value of the magnetic
field strength when the effect of the inclination and filling factor are considered. This MMAP value is
very robust to the specific hyperparameters of the prior $p(B)$.

\begin{figure*}
\centering
\includegraphics[width=\textwidth]{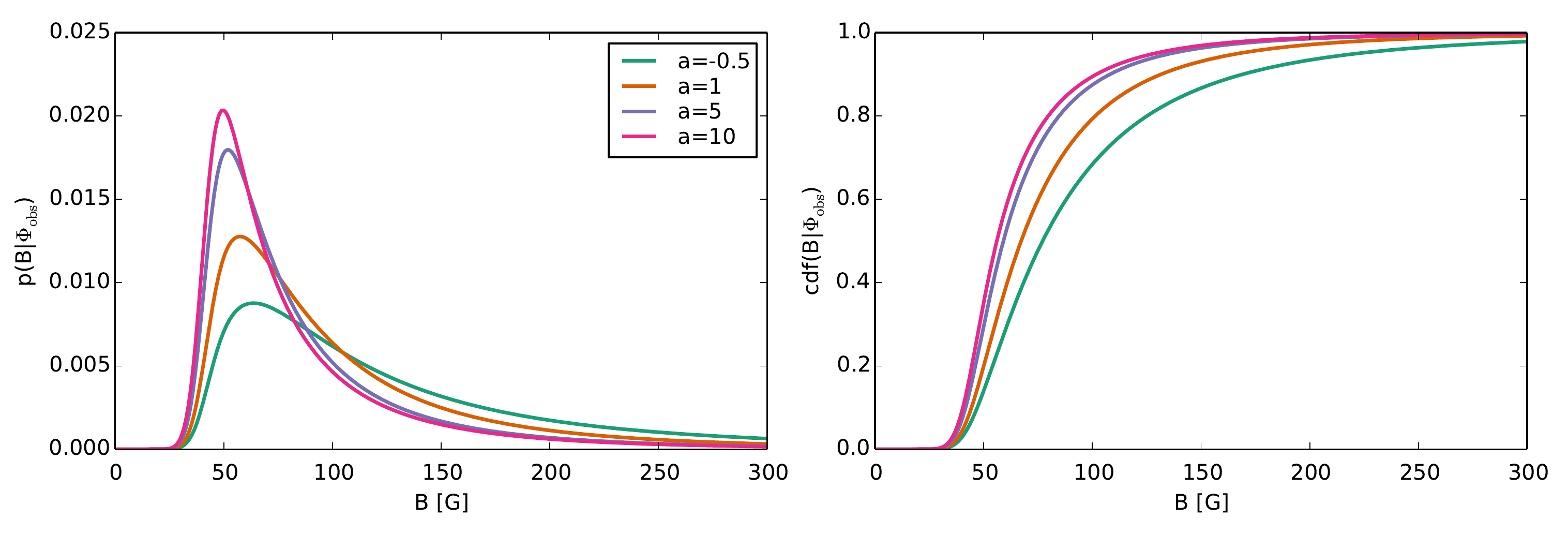}
\caption{Same as Fig. \ref{fig:posteriorFlux} for $\sigma_n=5$ Mx cm$^{-2}$ and $\Phi_\mathrm{obs}=40$ Mx cm$^{-2}$ and
different values of the anisotropic index of the prior for $\mu$.}
\label{fig:posteriorAnisot}
\end{figure*}

A problem that arises with the MMAP value is that it is biased towards small values of the field strength, given that the posterior
distribution is heavily skewed. More information can be extracted, though, from the cumulative distribution functions, shown in the right panels of Figs. \ref{fig:posteriorFlux}
and \ref{fig:posteriorNoise}, because they allow us to put strict upper limits to the magnetic field. From these curves, we have extracted the
percentiles 68 and 95 and they are shown in the second and third panel of Fig. \ref{fig:calibration}. Again, when
the noise is not too large, they approximately fulfill the power-laws shown in each panel.
These percentiles are also quite robust to the specific election of the prior and its hyperparameters, although
slighly more sensitive than the MMAP value.

\begin{figure*}
\centering
\includegraphics[width=\textwidth]{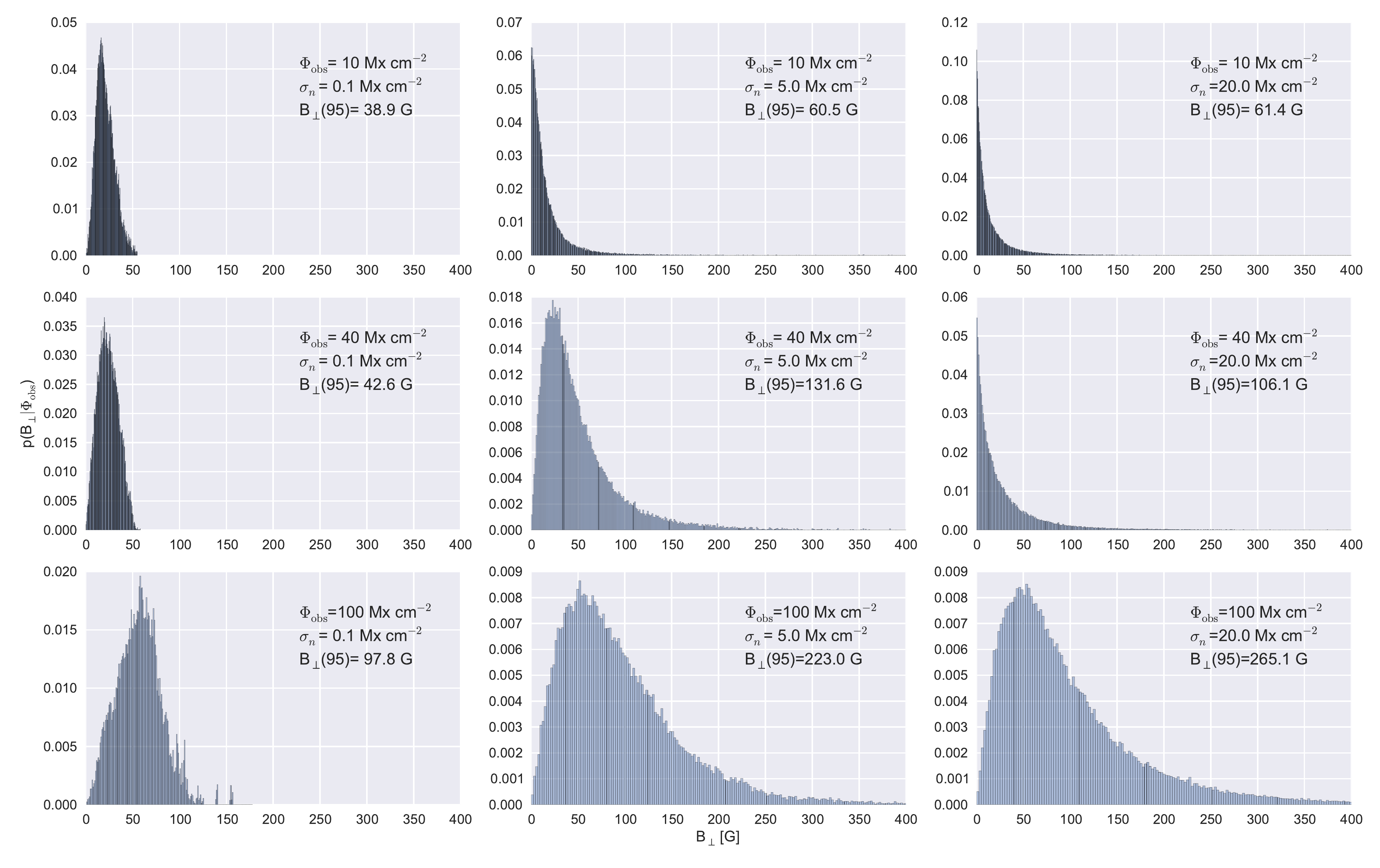}
\caption{Posterior distribution for the transversal component of the field for different values of $\Phi_\mathrm{obs}$
and $\sigma_n$. The value of the upper limit transverse field at 95\% credibility is shown in each panel. The bin
size is optimally chosen for each case.}
\label{fig:posteriorTrasverse}
\end{figure*}

\subsection{Anisotropic field distributions}
It has been argued in the literature, both observationally \citep{stenflo10} and from
simulations \citep{schussler08}, that the distribution of fields might be anisotropic. A simple
way of imposing this a-priori is by modifying the flat prior for $\mu$ to
\begin{equation}
p(\mu) \propto |\mu|^a,
\end{equation}
where the parameter $a$ controls the anisotropy ($a=0$ is an isotropic distribution, $a \to \infty$ corresponds
to vectors aligned with the LOS, while $a<0$ force fields to be predominantly orthogonal to the LOS). Figure
\ref{fig:posteriorAnisot} displays the marginal posterior for different values of $a$ for the log-normal prior. For increasing values of
$a$, the tails of the marginal posterior become less and less important, converging in the limit $a \to \infty$ to a Gaussian distribution
with mean $\Phi_\mathrm{obs}$ and variance $\sigma_n^2$, i.e., what one would obtain
by carrying out the inference assuming that $|\mu|=1$. For not very anisotropic distributions, the MMAP value is very robust to the specific
anisotropy of the field distribution, so that the corrections derived above can be reliably applied.

\subsection{Transversal component}
The Bayesian analysis we have carried out also allows us to put some constraints on the transversal component of the magnetic 
field, $B_\perp$. This might
look surprising but it is obvious that putting limits to $B$ automatically puts limits on $B_\perp$. Give that the transversal component
is given as a change of variables by $B_\perp=B f \sqrt{1-\mu^2}$, it is not possible to do much advance analitically. For this reason,
we consider a numerical solution to the problem by sampling from the full posterior of Eq. (\ref{eq:fullPosterior}) using
a Markov Chain Monte Carlo method \footnote{We use the \texttt{emcee} package developed by \cite{emcee12}.}
and compute the marginal posterior from the samples. The results are shown in Fig. \ref{fig:posteriorTrasverse}, where we consider
different values of the observed magnetic flux density and standard deviation of the noise. Each panel shows also the 
estimated value of the transversal component of the field at 95\% credibility.

It is obvious that the correct way of having an estimation of the transversal component of the field is by
observing the linear polarization components of the Stokes parameter. However, the results of Fig. \ref{fig:posteriorTrasverse}
provide an upper limit based on sensible assumptions and on the observation of the magnetic flux density alone.
When linear polarization is available, it is necessary to augment the generative model to include this information, 
as done by \cite{asensio_bayeswf11}.

\begin{figure*}
\centering
\includegraphics[width=\textwidth]{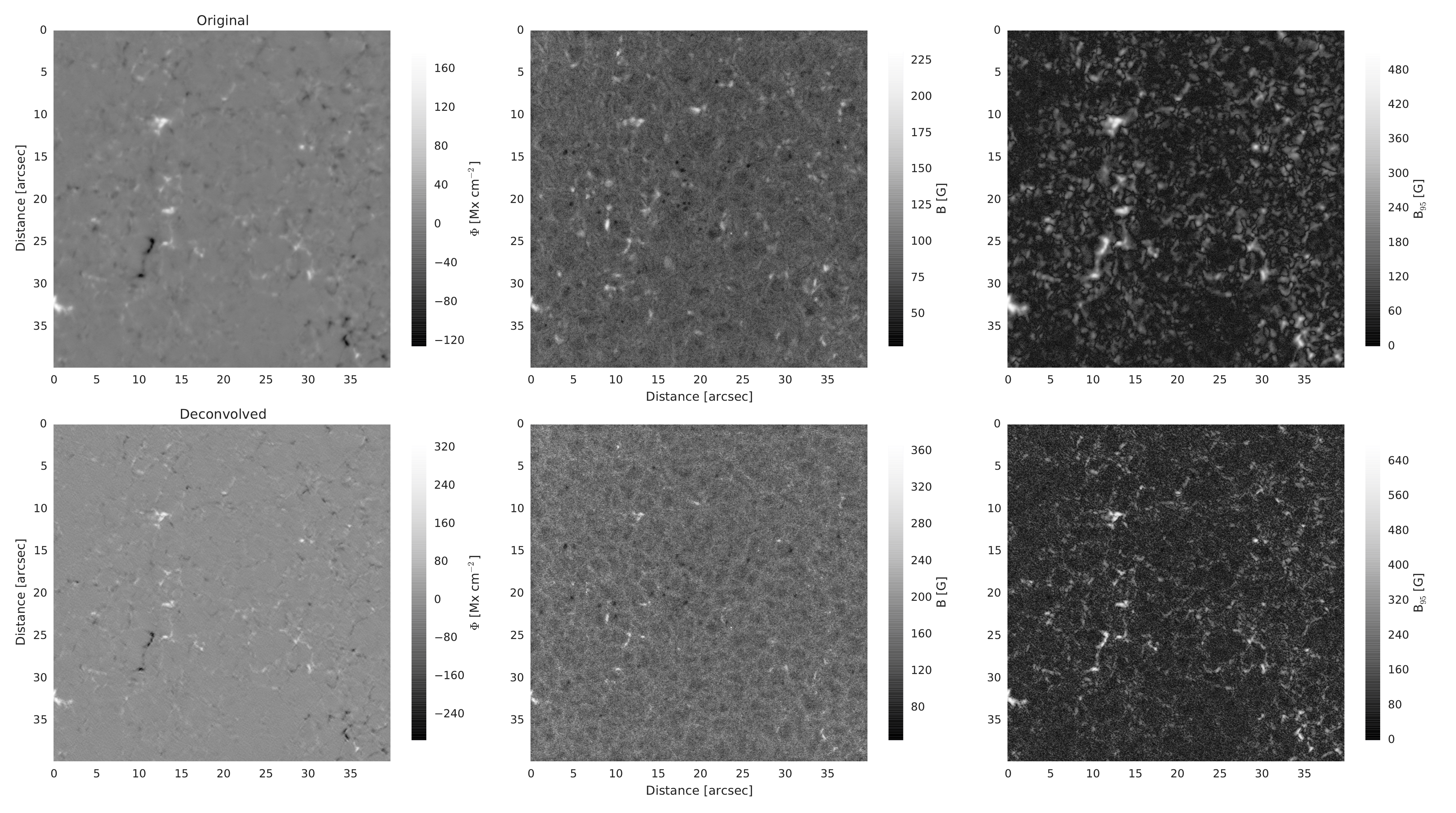}
\caption{Original (upper panels) and deconvolved (lower panels) inferred parameters
for the IMaX data of the first flight. The first column shows the magnetic flux
density. The middle column displays the inferred magnetic field strength, obtained
as $(B_\parallel^2+B_\perp^2)^{1/2}$. The last column displays our inferred upper
limit to the field strength at 95\% credibility.}
\label{fig:imax}
\end{figure*}

\begin{figure}[!b]
\centering
\includegraphics[width=0.78\columnwidth]{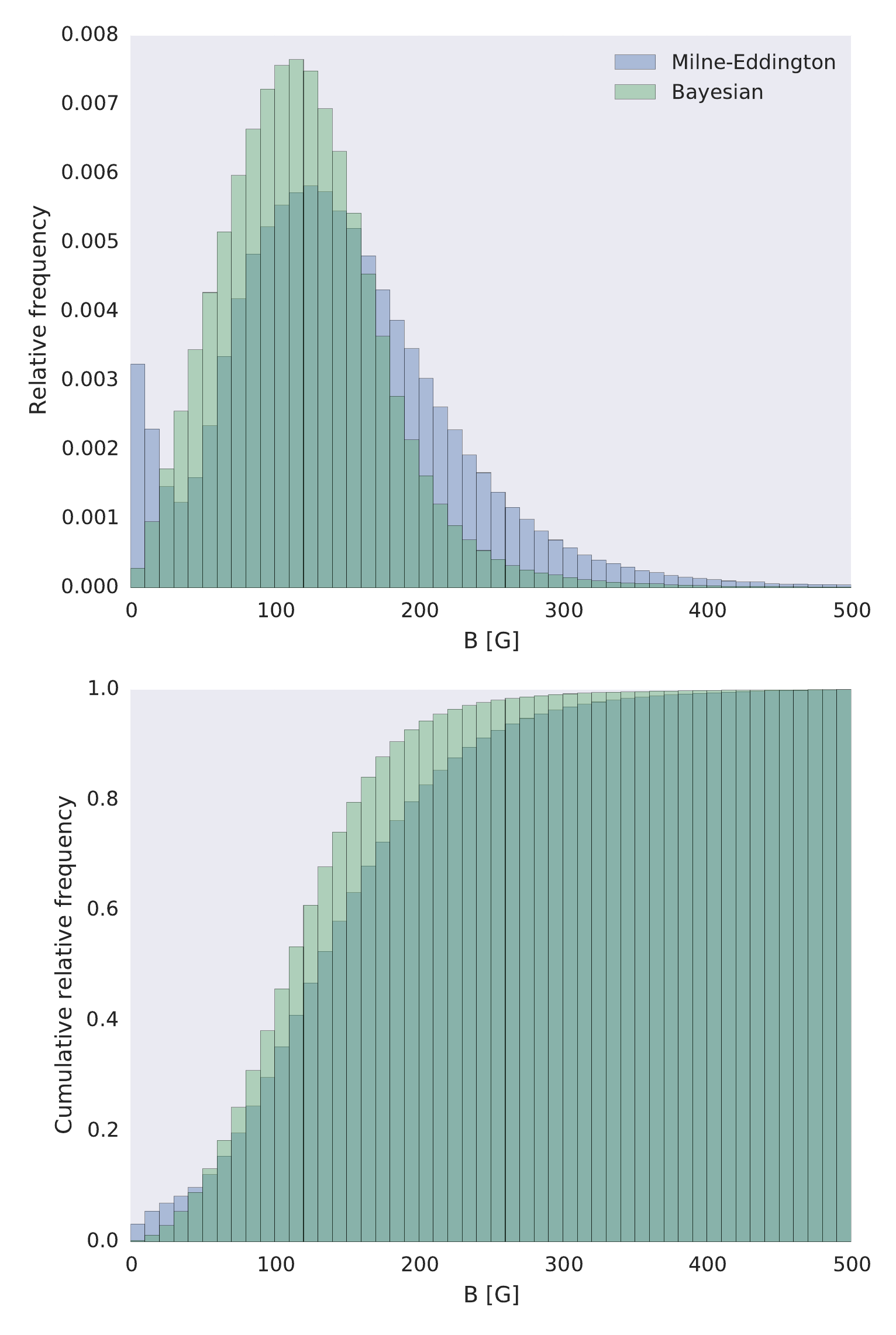}
\caption{Comparison of the histogram (upper panel) and cumulative histogram (lower panel) of the magnetic field
strength computed using a standard inversion code and our Bayesian approach at 95\% credibility.}
\label{fig:imaxHistogram}
\end{figure}

\subsection{Application to real data}
We apply the previous formalism to the data obtained with IMaX \citep{imax11} onboard Sunrise 
\citep{sunrise10} on the first flight on June 2009. Although IMaX is a vectorial magnetograph, that 
measures the four Stokes parameters, we only make use of the inferred magnetic flux density 
$\Phi_\mathrm{obs}$ in our estimation of the magnetic field strength. One of the particularities of the IMaX
mission is that both the original maps and those reconstructed with the phase-diversity 
post-facto reconstruction algorithm were provided \citep{paxman92,santiago_vargas09}. The left column
of Fig. \ref{fig:imax}
displays the maps of $\Phi_\mathrm{obs}$ in the two cases, demonstrating that the magnetic structures
in the reconstructed maps are more compact than in the original ones. To have a rough estimation of 
the magnetic field strength, we compute it as $(B_\parallel^2+B_\perp^2)^{1/2}$, where $B_\parallel$ and
$B_\perp$ are obtained using the simple calibration curves of \cite{imax11}. This is an imprecise but probably
conservative approximation for a filter-based instrument. This estimation is displayed in the middle column. As 
noted by \cite{marian_dipolo12}, the maximum-likelihood estimation of $B_\perp$
is biased in the presence of noise (because $B_\perp$ is obtained from $(Q^2+U^2)^{1/2}$, a positive definite 
quantity even in the absence of signal), which results in the appearance of a pattern that mimics inverse granulation.
Finally, the right columns of Fig. \ref{fig:imax} display the upper limit at 95\% credibility
using our estimations for an exponential prior with $B_0$=130 G and a very small
noise variance (see the lowest central panel of Fig. \ref{fig:calibration}). The upper limit field is larger for the deconvolved data, a consequence of correcting
for the spatial spread introduced by the instrument.

As a further check of our approach, we have compared the upper limit at 95\% credibility for the
magnetic field strength with the results of a standard inversion with the SIR code 
\citep[Stokes Inversion based on Response function;][]{sir92} of the data (kindly provided by L. Bellot Rubio) 
using a simple model comparable to a Milne-Eddington atmosphere. Figure \ref{fig:imaxHistogram} displays
the results in the form of histograms (upper panel) and cumulative histograms (lower panel). The results are 
very similar, even reproducing the position of the peak.

\section{Conclusions}
We have shown that the application of Bayesian inference to the data obtained with
longitudinal magnetographs can give some information about the magnetic field strength.
This information is a product of the presence of priors on the magnetic filling factor and
the inclination of the field. Our calculations have been done with very uninformative priors on
the angular distribution of the magnetic field and prior distributions for the magnetic field strength extracted from
analysis of simulations and observations of the quiet Sun. We demonstrate that our results are relatively robust
to the specific election of the prior.

The calculations can be repeated introducing more informative priors for $f$, $\mu$ and $B$ in Eq. (\ref{eq:posterior}).
Irrespectively of the chosen prior, Eq. (\ref{eq:posterior}) is easy to compute using any suitable numerical
quadrature for two-dimensional integrals.

\begin{acknowledgements}
We thank I. Arregui for very insightful discussions and L. Bellot Rubio for providing the quantities used in Fig. \ref{fig:imaxHistogram}.
Financial support by the Spanish Ministry of Economy and Competitiveness 
through projects AYA2010--18029 (Solar Magnetism and Astrophysical Spectropolarimetry) and Consolider-Ingenio 2010 CSD2009-00038 
are gratefully acknowledged. AAR also acknowledges financial support through the Ram\'on y Cajal fellowships.

\end{acknowledgements}

% \bibliographystyle{aa}
% \bibliography{/scratch/Dropbox/biblio}

\end{document}